\documentclass{aastex6}

\fullcollaborationName{The Friends of AASTeX Collaboration}
\newcommand{\beq}{\begin{equation}}
\newcommand{\eeq}{\end{equation}}
\newcommand{\bea}{\begin{eqnarray}}
\newcommand{\eea}{\end{eqnarray}}
\def\cor#1{\langle #1 \rangle}
\def\hsp{,\hspace{.7cm}}

\begin{document}


\title{A Robust Measure of Dark Matter Halo Ellipticities}

\author{Jarah Evslin}
\affiliation{Institute of Modern Physics, CAS, NanChangLu 509, Lanzhou 730000, China
}
\begin{abstract}
In simulations of the standard cosmological model ($\Lambda$CDM), dark matter halos are aspherical.  However, so far the asphericity of an individual galaxy's halo has never been robustly established.  We use the Jeans equations to define a quantity which robustly characterizes a deviation from rotational symmetry.  This quantity is essentially the gravitational torque and it roughly provides the ellipticity projected along the line of sight.  We show that the Thirty Meter Telescope (TMT), with a single epoch of observations combined with those of the Gaia space telescope, can distinguish the $\Lambda$CDM value of the torque from zero for each Sculptor-like dwarf galaxy with a confidence between 0 and 5$\sigma$, depending on the orientation of each halo.  With two epochs of observations, TMT will achieve a 5$\sigma$ discovery of torque and so asphericity for most such galaxies, and so will provide a new and powerful test of the $\Lambda$CDM model.

\end{abstract}




\section{Introduction} \label{sec:intro}

Cold dark matter simulations agree that dark matter halos \citep{triaxaq,triaxmil} and their subhalos \citep{darksub} are not spherical, but instead constant density surfaces are roughly triaxial ellipsoids whose axes vary from surface to surface.  These shapes may be probed most cleanly in dwarf spheroidal (dSph) galaxies, the nearest of which are hosted by subhalos of the Milky Way's dark matter halo, as these often consist of 99\% dark matter or more.  Many dSphs have multiple stellar populations which themselves are triaxial, with axes which are distinct from each other \citep{fornax2ell} and therefore likely from that of the dark matter halo.  On the other hand, more exotic dark matter models, including many self-interacting dark matter models and Bose Einstein Condensate models, often have more spherical dark matter halo profiles \citep{sfer}, although quantifying this effect can be difficult \citep{kapsfer}.  Therefore a measurement of the asphericity would provide a critical test of the cold dark matter paradigm.

So far, the most robust determinations of halo shapes have been obtained for galaxy clusters.  Such determinations are quite complicated, requiring a combination of X-ray, strong lensing and Sunyaev-Zeldovich effect observations, as in \citet{triaxcluster}.  There is also a large literature which attempts to determine the shape of the Milky Way's own dark matter halo.  However different methods have led to differing results, as is reviewed for example in \citet{triaxaq}.

In the clean case of the dark matter subhalos inhabited by Milky Way satellites, the state of the art subhalo shape determination is \citet{triaxhayashi}.  In this paper the authors found that the halo shapes are far more aspherical than those in CDM simulations, thus suggesting strong tension with the CDM paradigm.  However, this paper makes several very strong assumptions, including an alignment of the stellar and dark matter halos, that the halos have two equal semi-principal axes and, even more critically, an assumption on the anisotropy of the stellar velocities which, in the spherical case, reduces to the assumption that the stellar velocities are isotropic.  A shift in the stellar velocity anisotropy has, as the authors themselves described, a similar effect on the observables as a shift in the halo ellipticity and so this assumption alone may be responsible for the conclusions of the study.  Unfortunately, the assumption cannot be lifted because the anisotropy itself cannot be determined using the line of sight stellar motions together with the Jeans equations \citep{degen}.

However this situation is about to change.  Right now the Gaia space telescope is measuring not only the line of sight velocities of many stars in each of these satellite galaxies, but for the first time it is also measuring their transverse velocities.  With the full three-dimensional velocities in hand, this degeneracy will be broken and so more robust determinations of the halo shapes may be attempted.  The precisions of these proper motion determinations will improve dramatically with the completion of the Thirty Meter Telescope (TMT) \citep{tmtdsc}.  In particular, using the instrument IRIS, with a single view of such a galaxy one can obtain 20 $\mu$as precision relative positions of all of the stars bright enough for Gaia's astrometry.  Combining this with Gaia's observations of the relative positions, for example six years earlier, one obtains an improvement in the velocity measurements of roughly a factor of four \citep{megaia}.  With a second epoch of TMT observations several years later, one to two orders of magnitude more stars become available and the astrometric precision drops well below the stellar dispersions, which means that the measurements are limited primarily by the stellar dispersion and not the measurement error.

In summary, our knowledge of stellar proper motions will have three great leaps.  First, in two or three years when Gaia astrometry results become available, then when these results are combined with the first epoch of TMT observations and finally when the second epoch of TMT observations are combined with the first.  In the rest of this letter we will show that Gaia will be unable to meaningfully constrain the halo shapes, however the first epoch of TMT observations can distinguish spherical from triaxial CDM halos inhabited by Sculptor-like dwarf spheroidal (dSph) galaxies with up to $5\sigma$ of precision for a favorable orientation, although for a generic orientation the confidence is around 3$\sigma$.  However, since there are several such systems available, these 3$\sigma$ hints can be combined into a robust signal.  On the other hand, we will see that the second epoch of TMT observations will provide an overwhelming and robust signal which excludes, at the 5$\sigma$ level, either CDM or else spherical halos.

\section{Jeans Equation for the Gravitational Torque}

We will treat the stars in a galaxy as a collisionless gas in a potential $V({\bf{r}})$ with phase space density $f({\bf x},{\bf v})$, where ${\bf{x}}$ and ${\bf{v}}$ are the position and velocity three-vectors.  Then Boltzmann's equations are
\beq
0=\frac{df}{dt}=\frac{\partial f}{\partial t} + v_i \frac{\partial f}{\partial x_i}-\frac{\partial V}{\partial x_i}\frac{\partial f}{\partial v_i} \label{lioueq}
\eeq
where the first equality is Liouville's theorem and the second is the chain rule.  We will adapt cylindrical coordinates $(z,\rho,\theta)$ where $z$ is the line of sight direction, so that $\rho v_\theta={\rm{sin}}(\theta)v_1-{\rm{cos}}(\theta)v_2$.  The $\rho$ on the left results from the fact that we have defined $v_\theta=\partial\theta/\partial t$ to be the angular velocity, not the velocity in the $\theta$ direction.  Therefore $v_\theta$ has units of inverse time.

We will make two approximations.  First, we neglect the contribution of the baryons to the gravitational potential~$V$.  This is a reasonable assumption everywhere except for the interiors of the half-light radii of the most luminous dSphs, such as Fornax and Sculptor, where one may expect errors of order 10\%.  This assumption is easily removed with the understanding that the potential traced results from the sum of that of the dark matter and the stars, and so the results of our analysis apply to the total distribution.  In this case, to determine the dark matter distribution one needs to subtract the stellar distribution.  This can be done fairly precisely as only the line of sight integrated stellar mass density is needed.  Second, and more seriously, we assume that our configuration is in equilibrium and so $\partial f/\partial t=0$.   In contrast with the previous assumption, this is most reliable in the inner parts of the target galaxies, and in more distant galaxies.  In general equilibrium may be tested by evaluating spherical harmonics of odd moments of the radial velocity.  This may even be done with the projected radial velocity, which will be available at these surveys.

Then multiplying Eq.~(\ref{lioueq}) by $v_\theta$ and integrating over ${\bf{v}}$ one obtains the angular Jeans equation
\bea
\eta_s\partial_\theta V&=&-\frac{1}{\rho}\partial_\rho\left(\rho^3\eta_s\cor{v_\rho v_\theta}\right)-\rho^2\partial_\theta\left(\eta_s\cor{v_\theta^2}\right)\nonumber\\&&+\rho^2\partial_z\left(\eta_s\cor{v_\theta v_z}\right) \label{jeaneq}
\eea
where $\eta_s({\bf{x}})$ is the stellar luminosity density in 3-dimensional space, defined so that
\beq
\int d^3{\bf{v}} f(x,v) v_i v_j=\eta_s(x)\cor{v_iv_j}(x). 
\eeq
The ${\bf{x}}$-dependence of the moments $\cor{v_iv_j}$ and of the luminosity density $\eta_s$ will be left implicit.

As is, the Jeans equation (\ref{jeaneq}) may look useless not only because the full 3-dimensional dependences of $\eta_s$ and the moments are unknown, but even more seriously because the relative $z$-coordinates of individual stars cannot be measured in other galaxies given the astrometric precisions that will be available in the foreseeable future, and so $\partial_z$ is entirely unconstrained.

The good news is that all of these problems have the same solution.  To re-express the Jeans equation in terms of observable quantities, one needs to integrate Eq.~(\ref{jeaneq}) over the line of sight direction $z$, yielding
\bea
\int dz\left(\eta_s\partial_\theta V\right)&=&-\frac{1}{\rho}\partial_\rho\left(\rho^3\eta_s^{(2d)}\cor{v_\rho v_\theta}_{2d}\right)\nonumber\\&&-\rho^2\partial_\theta\left(\eta_s^{(2d)}\cor{v_\theta^2}_{2d}\right) \label{jean2d}
\eea
where we have defined the 2-dimensional luminosity density and moments, which implicitly depend on $(\rho,\theta)$, by
\beq
\eta_s^{(2d)}\cor{v_iv_j}_{2d}=\int dz \eta_s \cor{v_i v_j}.
\eeq
Here the unknowable $\partial_z$ term in Eq.~(\ref{jeaneq}) has vanished as a result of the fundamental theorem of calculus and the fact that $\eta_s$ vanishes at large positive and negative values of $z$.

The disappearance of the $\partial_z$ term, via integration by parts, is the central observation behind this paper.  It implies that no assumptions need to be made concerning the deprojection of the image.  There is no such disappearance for the other Jeans equations, and so the results that follow may only be applied to the angular Jeans equation (\ref{jeaneq}).  In general, given proper motion data, orbit-based methods yield more robust results than Jeans equations-based methods because the Jeans equations do not imply the dynamical consistency of the system, and so additional assumptions need to be invoked.  However the case at hand provides a counter-example to this common wisdom, as orbit-based methods need to make additional assumptions regarding the deprojection and the angular Jeans equation does not.  Of course, the price to pay for this robustness is that only the angular Jeans equation is available.  The angular Jeans equation is sufficient to obtain some halo properties such as the projected ellipticity.   However other properties, such as the radial density profile, cannot be determined from the angular Jeans equation alone and so orbit-based methods remain the most powerful for such goals.

The left-hand side of Eq.~(\ref{jean2d}) is a luminosity-weighted expression for the angular dependence of the gravitational potential.  More precisely, it is $(1/\rho)$ times the gravitational torque which is exerted on all of the stars in the line of sight at fixed $(\rho,\theta)$. For a given stellar distribution and dark matter density profile, it may be calculated.  The right-hand side, on the other hand, consists entirely of quantities which may be measured.  Thus the strategy will be to measure the right hand side so as to determine the left hand side.

The left hand side has one very attractive feature.  It vanishes for a spherically symmetric mass density, with no restriction on the stellar luminosity profile or velocity anisotropy.  Therefore if a measurement of the right hand side in a strongly dark matter dominated system shows that this quantity is nonzero, it cannot be attributed entirely to the unknown 3-d stellar distribution and velocity anisotropy or to the halo's unknown radial density profile, it implies robustly that the dark matter halo itself is indeed not spherical.  Thus Eq.~(\ref{jean2d}) provides a very clean test for the asphericity of an individual dark matter halo.  {\it The fact that the torque defined by the left hand side of Eq.~(\ref{jean2d}) provides an unbiased and robust statistic for testing the asphericity of the matter distribution is our main result.}

\section{Precision with which torque can be measured}

In the rest of this letter, we will determine the confidence with which Gaia space telescope and Thirty Meter Telescope (TMT) observations, using the instrument IRIS, can, using the quantity defined in Eq.~(\ref{jean2d}), distinguish a spherical dark matter halo inhabited by a galaxy similar to the Sculptor dwarf spheroidal from a triaxial dark matter halo with axis ratios of the order seen in CDM simulations.  For simplicity, we will assume that the total matter density follows an NFW profile
\beq
\rho(R)=\frac{\rho_0 R_0^2}{R(R+R_0)^2} \label{nfw}
\eeq
where $r_0=0.5$ kpc and $\rho_0=8\times 10^7\ M_\odot/{\rm{kpc}}^3$ corresponding to the best fit values for the Sculptor dwarf in \citet{strig2006}.  Here $R$ is defined by
\beq
R^2=\frac{x_1^2}{a_1^2}+\frac{x_2^2}{a_2^2}+\frac{x_3^2}{a_3^2} \label{rdef}
\eeq
where in the spherically symmetric case $a_i$=1 and in the triaxial case we will choose
\beq
a_1=1\hsp a_2=0.6\hsp a_3=0.9 \label{asse}
\eeq
reflecting typical axis ratios found in $\Lambda$CDM simulations.  The corresponding gravitational potentials are evaluated numerically, by summing the potentials created by constant density ellipsoids.

\begin{figure} %[!tph]
\begin{center}
\includegraphics[width=2.8in,height=1.1in]{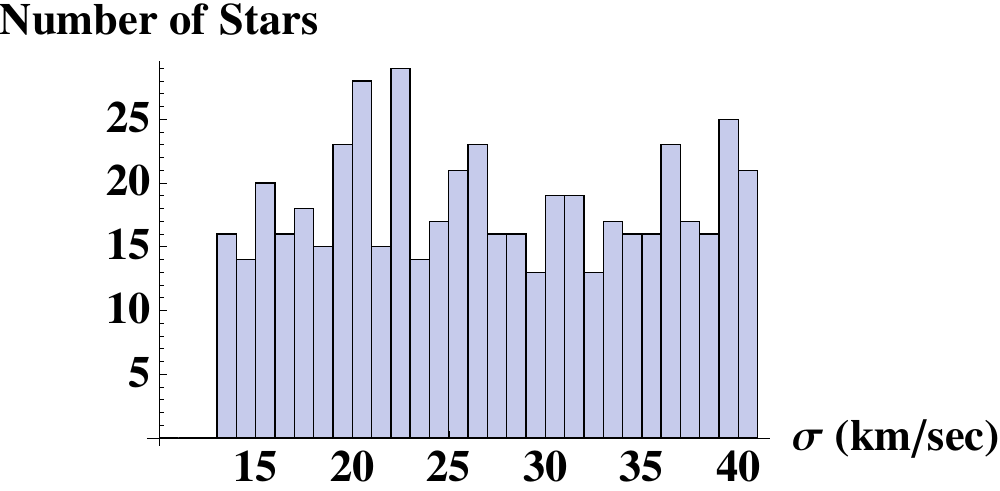}
%\plotone{1epoch.pdf}
\caption{The number of Sculptor dwarf members whose proper motion the Gaia satellite can measure with a given precision $\sigma$.  Only the 1355 members in the MMFS survey are considered, and so these numbers are lower bounds.  TMT uncertainties for observations of the same stars in 2022 can be obtained by dividing $\sigma$ by 4.}
\label{gaiafig}
\end{center}
\end{figure}

Although TMT can offer precise astrometry for well over $10^4$ stars in the Sculptor dwarf \citep{tmtdsc}, we will restrict our attention to the 1355 stars in the MMFS survey \citep{mmfs} for which the authors have assigned a membership probability greater than 90\%.  The uncertainties with which Gaia can determine proper motions are determined using the post-launch precision study in Ref.~\citep{gaia2015} with a 15\% improvement due to Sculptor's favorable location with respect to Gaia's observation pattern.  Following \citet{megaia}, in Fig.~\ref{gaiafig} we present the number of MMFS Sculptor members whose proper motions can be measured with various uncertainties.  These uncertainties are always greater than the measured radial velocity dispersion, however the large number of stars means that they nonetheless provide a reasonably accurate measurement of the 3d velocity dispersion profile.

As these stars all have $H$ band magnitudes well below 21, TMT's astrometric precision will be limited by systematic errors, which we will conservatively approximate to be 20 $\mu$as.  As a result, as shown in \citet{megaia}, a single epoch of TMT observations combined with Gaia, 6 years earlier, yields an astrometric precision which is about 4 times better than Gaia alone.  We also consider a second epoch of TMT observations, which yields a 2 km/s precision on each of these stars.  If instead of the TMT observations with IRIS we consider the European Extremely Large Telescope (E-ELT) astrometry instrument ELT-CAM, then two epochs separated by a 5 year baseline are required to yield a 5 km/s astrometric precision.

Can IRIS with NFIRAOS adaptive optics observe all of these stars?  No.  The science requirement for NFIRAOS is that there should be enough guidestars to cover at least half of the potential fields of view at the galactic poles, and more elsewhere.  We have not considered this reduction in the number of stars in our analysis, as the size of this reduction is not yet known beyond the fact that it is beneath 50\%.

For the remaining stars, these observations are possible as a result of the redesign of the IRIS instrument to have a field of view of 35$''\times$35$''$.   As a result, less than 1200 fields are sufficient to map Sculptor out to its half light radius of 11.3$'$ \citep{mcconnachie}.   For the fields observed, using the TMT exposure time calculator one finds that a 60 second observation per epoch yields the required 20 $\mu$as (50 $\mu$as) astrometric precision down to magnitude H=21 (22).  At H$\lesssim$21, there are about 10 Sculptor members per field inside of the half-light radius \citep{deboer}, and more than 100 within the field of view of NFIRAOS, which will be scanned and pasted together.  Assuming that IRIS can stably mosaic these fields together, this will be quite sufficient for differential astrometry.  In addition, each IRIS field will typically contain several members of the Gaia catalog, increasing further the precision.  In more sparsely populated fields, one may use the substructure of distant galaxies to fix a frame \citep{distgal}.  To cover the 1355 MMFS Sculptor members or the roughly $10^4$ members \citep{deboer} in the 1200 fields within the half-light radius, with 60 second exposures, one needs about 20 hours of observation per epoch with IRIS, and probably less once WISC is operational.  The TMT Key Project on Local Group dwarf galaxies envisages significantly more observing time, allowing deeper and wider observations.

To determine the precision with which a given set of observations may distinguish a spherical dark matter halo from a triaxial dark matter halo, we define a $\chi^2$ statistic to be
\beq
\chi^2=\sum_{\rm stars}\frac{\left(\int dz\left(\eta_s\partial_\theta V\right)\right)^2}{{\rm var}(\rho,\sigma)}
\eeq
where the numerator and denominator are evaluated at the location of each star in the sum and the variance for a given star is
\bea
{\rm{var}}(\rho,\sigma)&=&\frac{8\rho^4\eta_s^{(2d)\ 2}}{\pi^2}\left(\sigma^2+\langle v_\theta^2\rangle_{\rm 2d}\right)^2\\
&+&\left[\left(\rho^2\frac{\partial\eta_s^{(2d)}}{\partial\rho}+3\rho\eta_s^{(2d)}\right)^2+\frac{\rho^4}{\rho_m^2}\eta_s^{(2d)\ 2}\right]\nonumber\\&&\times\left(\sigma^2+\langle v_\rho^2\rangle_{\rm 2d}\right)\left(\sigma^2+\langle v_\theta^2\rangle_{\rm 2d}\right).\nonumber
\eea
Here $\rho_m$ is chosen to be 0.23 kpc, the projected half-light radius, to impose the crude approximation that the uncertainty in a measurement of $\partial_\rho X$ is equal to $1/\rho_m$ times the uncertainty in $X$.  Similarly, the uncertainty in $\partial_\theta X$ is taken to be the uncertainty in $X$ divided by $\pi/2$, as the angular dependence of $V$ is dominated by the quadrupole moment for a triaxial dark matter distribution.  Here $\sigma$ is the proper motion measurement uncertainty.  The velocity moments are calculated from the dark matter distribution (\ref{nfw}) by integrating the second order Jeans equation assuming spherical symmetry and also the isotropy condition $\langle v_r^2\rangle=\langle v_\phi^2\rangle$. 

We will see that the orientation of the triaxial dark matter halo strongly affects the confidence with which the torque or asphericity can be measured.  Recall that the triaxiality was incorporated into our halo density profile by the definition (\ref{rdef}) in which we chose the semi-principal axes (\ref{asse}), so that the profile in the $x_1-x_2$ plane is the most elliptical, while the projection to the $x_1-x_3$ plane is the roundest.  We relate the triaxial coordinates ${\bf x}$ to the observational coordinates $(\rho,\theta,z)$ by the matrix
\bea
\left(
\begin{tabular}{c}
  $x_1$\\$x_2$\\$x_3$
\end{tabular}
\right)
&=&
\left(
\begin{tabular}{ccc}
  ${\rm cos}(\Phi)$&${\rm sin}(\Phi)$&$0$\\
 -${\rm sin}(\Phi)$&${\rm cos}(\Phi)$&$0$\\
  $0$&$0$&$1$
\end{tabular}
\right)\\
&&\times\left(
\begin{tabular}{ccc}
  ${\rm cos}(\Theta)$&$0$&${\rm sin}(\Theta)$\\
y  $0$&$1$&$0$\\
 $-{\rm sin}(\Theta)$&$0$&${\rm cos}(\Theta)$\\
\end{tabular}
\right)
\left(
\begin{tabular}{c}
  $\rho{\rm sin}(\theta)$\\$\rho{\rm cos}(\theta)$\\$z$
\end{tabular}
\right)
\nonumber
\eea
so that the projected ellipticity depends on the two angles $\Theta$ and $\Phi$.  In particular, if $\Theta=\Phi=0^\circ$ then the visible $(\rho,\theta)$ plane corresponds to the $x_1-x_2$ plane and so the projected ellipticity is large, allowing for a clean signal, whereas $\Theta=\Phi=90^\circ$ implies that the $x_1-x_3$ plane is orthogonal to the line of sight, and so the projected ellipticity is small.  In practice the confidence with which the torque and so ellipticity can be distinguished from zero depends not only on the direction of the line of sight with respect to the axes, but also, to a lesser extent, on the distribution of stars.

\begin{figure} %[!tph]
\begin{center}
\includegraphics[width=2.8in,height=2.4in]{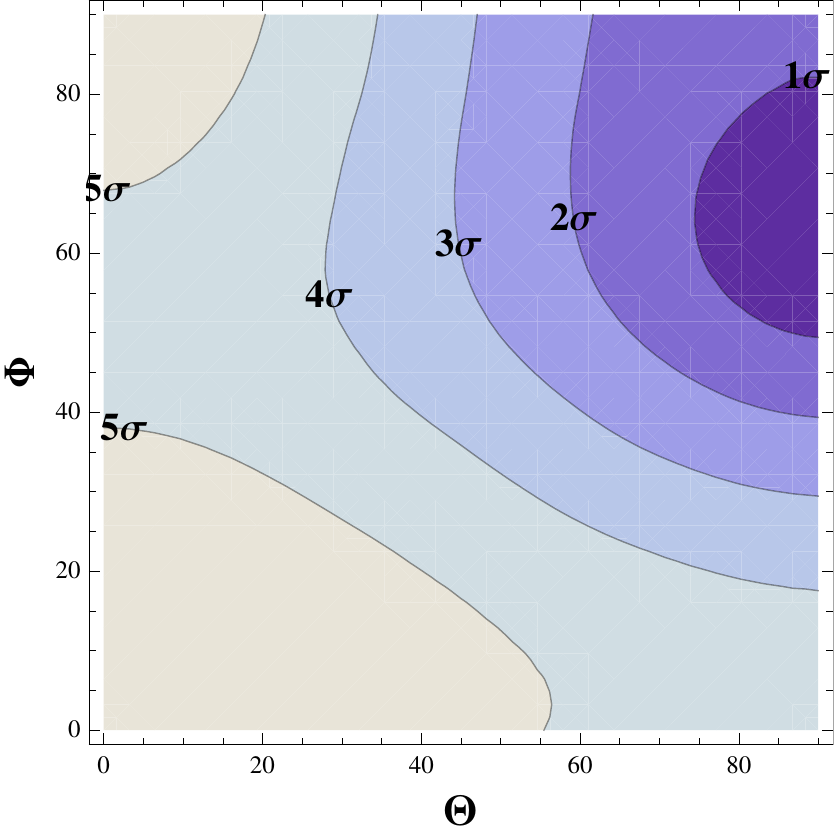}
%\plotone{1epoch.pdf}
\caption{The confidence with which the Gaia space telescope, combined with a single epoch of TMT observations, can distinguish a spherically-symmetric matter distribution in a Sculptor-like dwarf spheroidal galaxy from a triaxial dark matter distribution with semi-principal axis ratios similar to those found in $\Lambda$CDM satellites.  The angles $\Theta$ and $\Phi$ characterize the orientations of the semi-principal axes with respect to the line of sight.}
\label{ep1fig}
\end{center}
\end{figure}

In the case of Gaia-only observations of stellar proper motions, we have found that even a 1$\sigma$ asphericity signal cannot be achieved.  However, a single epoch of TMT observations combined with Gaia observations 6 years earlier leads to 3$\sigma$ evidence asphericity for most axis orientations $(\Theta,\Phi)$, with some orientations yielding 5$\sigma$, as is shown in Fig.~\ref{ep1fig}.  Note that by combining observations of several dSphs, a 5$\sigma$ discovery of asphericity can be attained and these deviations from asphericity can be compared with simulations to test $\Lambda$CDM.  

\begin{figure} %[!tph]
\begin{center}
\includegraphics[width=2.8in,height=2.4in]{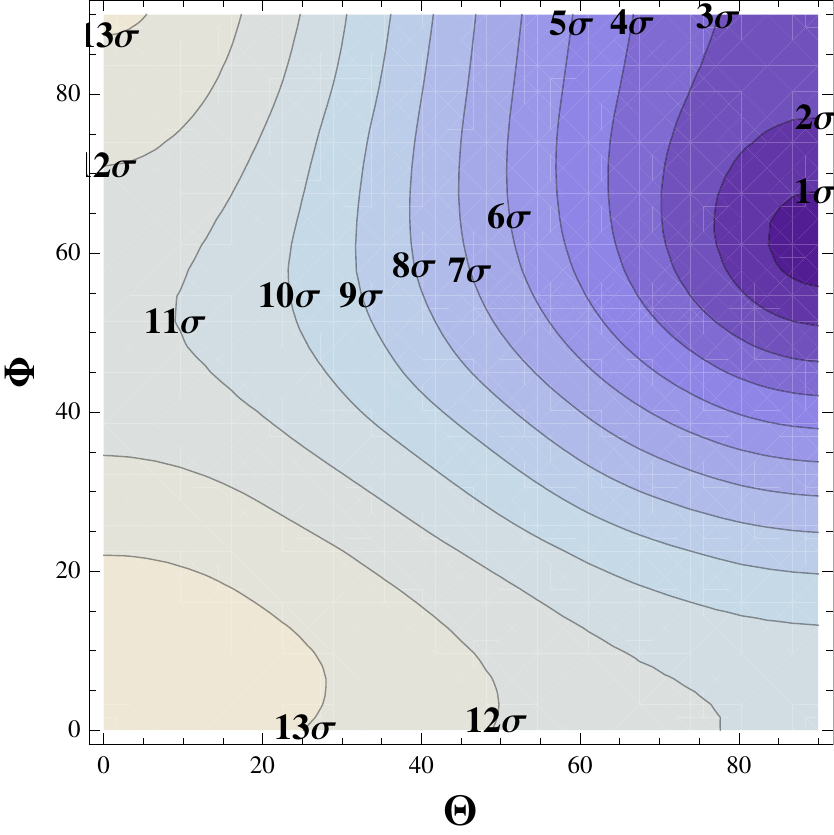}
\includegraphics[width=2.8in,height=2.4in]{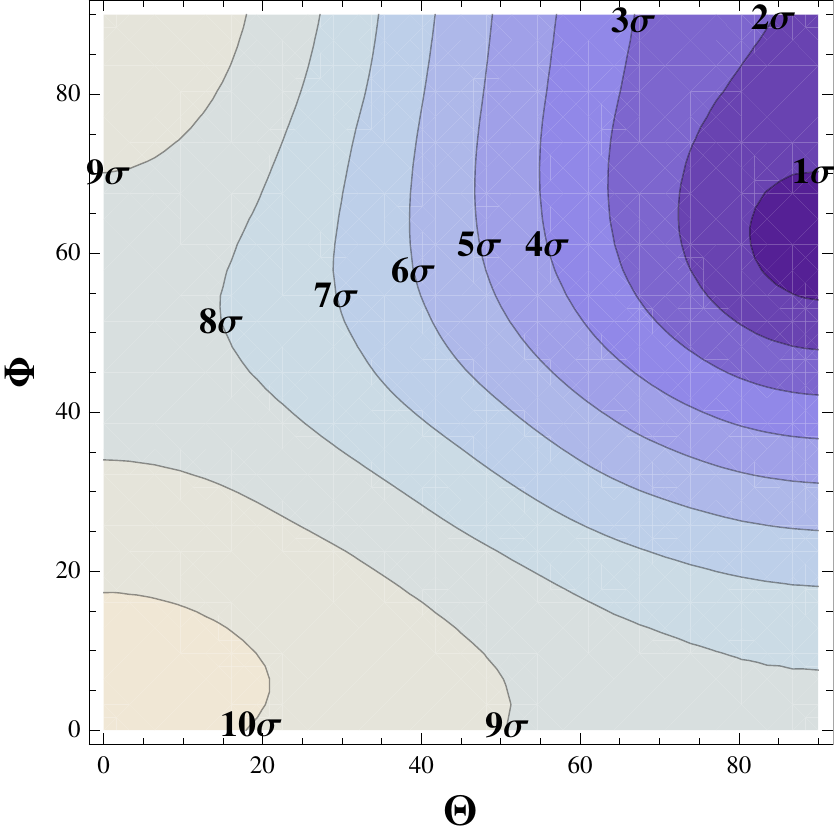}
\caption{As in Fig.~\ref{ep1fig} but for two epochs of TMT (left) or E-ELT (right) observations.}
\label{ep2fig}
\end{center}
\end{figure}

On the other hand, two epochs of either TMT or E-ELT observations yields a 5$\sigma$ discovery of asphericity for most orientations, as is shown in Fig.~\ref{ep2fig}.  As in the previous figure, when $\Theta=90^\circ$ and $\Phi\sim 60^\circ$, the confidence vanishes because the ellipsoid projected along the line of sight is circular. These confidences only consider the stars observed in the MMFS survey, however both TMT and E-ELT should be able to achieve comparable astrometry with an order of magnitude more stars, and so this figure is quite conservative.  In fact, these measurements will be so precise that we believe it will be possible to constrain even the radial dependence of the projected ellipticity, providing a very powerful test of $\Lambda$CDM.

\section* {Acknowledgement}

\noindent
I am supported by NSFC MianShang grant 11375201.   
%%%%%%%%%%%%%%%%%%%%%%%%%%%%%%%%


\begin{thebibliography}{99}%\setlength{\itemsep}{-2.3mm}

%%%%%%%%%%%%%%%%%%%%%%%%%%%%%%%%%


\bibitem[de Boer et al.(2011)]{deboer}
  de Boer, T.~J.~L., Tolstoy, E., Saha, A., et al.\ 2011, \aap, 528, A119 
  
\bibitem[de Bruijne et al. (2015)]{gaia2015}
de Bruijne, J. H. J., Rygl, K. L. J. and Antoja, T., 2015,  
``Gaia Astrometric Science Performance - Post-Launch Predictions," 
arXiv:1502.00791 [astro-ph.IM].


\bibitem[Dejonghe (1987)]{degen}
   Dejonghe, H., 1987,
  \mnras ,  {133}, 217.

\bibitem[Evslin (2015)]{megaia}
  Evslin, J.~, 2015,
  %``What can Gaia (with Thirty Meter Telescope) say about the Sculptor Dwarf's Core?,''
  \mnras ,  {452},  1,  L41
  %doi:10.1093/mnrasl/slv083
  %[arXiv:1501.07503 [astro-ph.GA]].

  
\bibitem[Hayashi and Chiba (2012)]{triaxhayashi}
  Hayashi, K. and Chiba, M., 2012,
  %``Probing non-spherical dark halos in the Galactic dwarf galaxies,''
  \apj ,  {755}, 145
  %doi:10.1088/0004-637X/755/2/145
  %[arXiv:1206.3888 [astro-ph.CO]].

\bibitem[McConnachie (2012)]{mcconnachie}  
McConnachie, A. W., 2012,
  %``The observed properties of dwarf galaxies in and around the Local Group,''
  \aj ,  {144}, 4
 % doi:10.1088/0004-6256/144/1/4
%  [arXiv:1204.1562 [astro-ph.CO]].  
  
\bibitem[Miralda-Scude (2002)]{sfer}
  Miralda-Escude, J., 2002,
  %``A test of the collisional dark matter hypothesis from cluster lensing,''
  \apj ,  { 564}, 60
  %[astro-ph/0002050].

 \bibitem[Morandi et al.(2012)]{triaxcluster}
 Morandi, A., Limousin, M., Sayers, J., Golwala, S.~R., Czakon, N.~G., Pierpaoli, E. and Ameglio, S., 2012,
  %``X-ray, lensing and Sunyaev Zel'dovich triaxial analysis of Abell 1835 out to R_{200},''
  \mnras , { 425}, 2069
 % doi:10.1111/j.1365-2966.2012.21196.x
  %[arXiv:1111.6189 [astro-ph.CO]]. 
 
  
\bibitem[Peter et al.(2013)]{kapsfer}
  Peter, A.~H.~G., Rocha, M., Bullock, J.~S. and Kaplinghat, M., 2013,
  %``Cosmological Simulations with Self-Interacting Dark Matter II: Halo Shapes vs. Observations,''
  \mnras , { 430}, 105
  %doi:10.1093/mnras/sts535
  %[arXiv:1208.3026 [astro-ph.CO]].

  
\bibitem[del Pino et al.(2015)]{fornax2ell}
  del Pino, A.,  Aparicio, A., Hidalgo, S. L., 2015,
  \mnras,  {454}, 3996
  %[arXiv:1509.05336  [astro-ph.GA]].
  
  
\bibitem[Schneider et al.(2012)]{triaxmil}
 Schneider, M.~D., Frenk, C.~S. and Cole, S., 2012,
  %``The Shapes and Alignments of Dark Matter Halos,''
  JCAP, {1205}, 030
%  doi:10.1088/1475-7516/2012/05/030
  %[arXiv:1111.5616 [astro-ph.CO]].

\bibitem[Skidmore et al. (2015)]{tmtdsc}
  Skidmore, W. {\it et al.} [TMT International Science Development Teams \& TMT Science Advisory Committee Collaboration], 2015,
  %``Thirty Meter Telescope Detailed Science Case: 2015,''
  Res.\ Astron.\ Astrophys.\ , { 15},  12,  1945
  %doi:10.1088/1674-4527/15/12/001
  %[arXiv:1505.01195 [astro-ph.IM]].

\bibitem[Strigari et al. (2007)]{strig2006}
  Strigari, L.~E.~, Koushiappas, S.~M.~, Bullock, J.~S. and Kaplinghat, M., 2007,
  %``Precise constraints on the dark matter content of Milky Way dwarf galaxies for gamma-ray experiments,''
%  Phys.\ Rev.\ D
 \prd , {75}, 083526
  %doi:10.1103/PhysRevD.75.083526
  %[astro-ph/0611925].

\bibitem[Trippe et al.(2010)]{distgal} Trippe, S., Davies, R., Eisenhauer, F., et al.\ 2010, \mnras, 402, 1126 
 
  
\bibitem[Vera-Ciro et al.(2011)]{triaxaq}
  Vera-Ciro, C.~A., Sales, L.~V., Helmi, A., Frenk, C.~S., Navarro, J.~F., Springel, V.~, Vogelsberger, M. and White, S.~D.~M.~, 2011,
  %``The Shape of Dark Matter Haloes in the Aquarius Simulations: Evolution and Memory,''
  \mnras, 416, 1377
%  Mon.\ Not.\ Roy.\ Astron.\ Soc.\  {\bf 416} (2011) 1377
%  doi:10.1111/j.1365-2966.2011.19134.x
%  [arXiv:1104.1566 [astro-ph.CO]].
  
  
  \bibitem[Vera-Ciro et al.(2014)]{darksub}
  Vera-Ciro, C., Sales, L.~V., Helmi, A. and Navarro, J.~F., 2014,
  %``The shape of dark matter subhaloes in the Aquarius simulations,''
  \mnras, {439}, 3,  2863
  %doi:10.1093/mnras/stu153
  %[arXiv:1402.0903 [astro-ph.CO]].
  



  



 
\bibitem[Walker et al. (2009)]{mmfs}
 Walker, M. G., Mateo, M. and Olszewski, E., 2009, 
%`Stellar Velocities in the Carina, Fornax, Sculptor and Sextans dSph Galaxies: Data from the Magellan/MMFS %Survey,''
 \aj , {137}, 3100
 %[arXiv:0811.0118 [astro-ph]]. 

 
\end{thebibliography}
\end{document}